\documentclass[pra,twocolumn]{revtex4}
\usepackage[english]{babel}
\usepackage{amsmath}
\usepackage{amssymb}
\usepackage{epsfig}
\usepackage{url}

\begin{document}
\title{Kelvin-Helmholtz instability in nonlinear optics}
\author{Victor P. Ruban}
\email{ruban@itp.ac.ru}
\affiliation{Landau Institute for Theoretical Physics RAS,
Chernogolovka, Moscow region, 142432 Russia}

\date{\today}

\begin{abstract}
Paraxial propagation of a quasi-monochromatic light 
wave with two circular polarizations in a defocusing 
Kerr medium with anomalous dispersion inside 
a waveguide of annular cross-section was considered. 
In the phase-separated mode, the dynamics 
is similar to a flow of immiscible fluids.
For some initial conditions with relative gliding of the 
fluids along the interface, the Kelvin-Helmholtz
instability in its ``quantum'' variant is developed. 
Numerical simulations of the corresponding coupled
nonlinear Schr\"odinger equations have shown formation 
of specific structures at the nonlinear stage
of the instability. Similar structures have been 
known in the theory of binary Bose-Einstein condensates,
but for optics they were presented for the first time.

\vspace{3mm}

\noindent {\bf DOI}: 10.31857/S00444510240214e3
\end{abstract}

\maketitle

\section{INTRODUCTION}

The study of nonlinear waves of different physical 
nature is, first of all, a search for possible coherent 
structures in a given system. Perhaps the best
known are such long-lived objects as solitons and
vortices (see, for example, [1-7]). Of great interest are
also structures formed as a result of the development
of different kinds of instabilities. A striking example
is the Kelvin-Helmholtz instability (KHI), which occurs 
in liquid systems in the presence of a tangential
discontinuity in the velocity field. A quantum analog
of this instability in superfluid 3 He has been discovered
relatively recently [8-11]. The KHI is also known for
two-component ultracold gases (Bose--Einstein condensates, 
BEC) in the spatial phase separation regime
[12-14], where it has been theoretically studied along
with other analogs of classical hydrodynamic instabilities [15-20]. 
The example of BSCs is particularly interesting 
because in the zero-temperature limit these
sparse systems are well described by a rather simple and
universal mathematical model -- the coupled nonlinear
Schr\"odinger equations [21-27] (in the physics of cold
Bose-gases they are called the Gross--Pitaevskii equations). 
In a binary BEC, in addition to dark solitons
and quantized vortices in each of the two components,
domain walls are still possible due to the dominance
of the crossed nonlinear interaction. The domain wall
has an effective surface tension [24, 28, 29], which 
significantly participates in the interface dynamics at all
stages of instability development.

The coupled nonlinear Schr\"odinger equations
(NSE) are known to arise in the study of many physical 
systems. In particular, they approximate the propagation 
of a quasi-monochromatic light wave with two
circular polarizations in an optical medium with Kerr
nonlinearity [30]. A possible difference with BECs
is the hyperbolic type of the three-dimensional dispersion 
operator in the case of normal optical dispersion.
But for anomalous dispersion and defocusing nonlinearity, 
the paraxial optical equations are mathematically 
equivalent to the two coupled Gross-Pitaevskii
equations in which the KHI is developed.
Moreover, when considering spatially two-dimensional
solutions, as in [12-14], the type of dispersion
is formally unimportant since there is no dependence
on the corresponding third coordinate (unless one
is interested in the dynamics of small three-dimensional 
perturbations).

It should be noted that in nonlinear optics domain
walls separating regions with right and left circular
polarizations of light are also known [31-40], but the
KHI of such interface surfaces in the presence of relative 
shear flow of ``light fluids'' has not been considered
so far. It should be emphasized that we are not talking
here about the primary modulation instability of spatially 
homogeneous two-component systems with
dominant cross-repulsion [41-43], which actually leads
to phase separation through the formation of domain
walls, but about the possible instability of the stationary 
geometrical configuration of the already existing
domain wall (analogous to the instability of dark solitons 
in one-component NSE [44, 45]). In this sense,
the KHI in the system of coupled NSEs can be called
secondary. The purpose of the present work is to numerically 
demonstrate the development of such instability 
in the framework of a model that approximates
the paraxial propagation of a nonlinear optical wave
in a sufficiently wide annular waveguide.

\section{MODEL DESCRIPTION}

Note at once that in optics the distance $\zeta$ along the waveguide 
axis serves as the evolutionary variable instead
of time $t$, and the role of the third ``spatial''
coordinate is played by the ``delayed'' time 
$$
\tau=t-\zeta/v_{\rm gr},
$$
where $v_{\rm gr}$ is the group velocity of light in the medium 
at a given carrier frequency $\omega_0$. When comparing
with BECs, one should mean $t$ instead of $\zeta$ and $z$ 
instead of $\tau$. In dimensionless variables, the nonlinear
Schr\"odinger equations for the slow complex envelopes
$A_{1,2}(x,y,\tau,\zeta)$ of the left and right circular polarizations
of the light wave are of the form (see [30-38])
\begin{equation}
i\frac{\partial A_{1,2}}{\partial \zeta}=\Big[-\frac{1}{2}\Delta +U(x,y)
+|A_{1,2}|^2+ g_{12}|A_{2,1}|^2\Big]A_{1,2},
\label{A_12_eqs}
\end{equation}
where
$
\Delta=\partial_x^2+\partial_y^2+\partial_\tau^2
$
is the three-dimensional Laplace operator in the ``coordinate'' 
space $(x,y,\tau)$. In the case of normal dispersion, 
the sign in front of $\partial_\tau^2$ is reversed. The potential
$U(x,y)$ is determined by the dielectric permittivity profile 
in the cross section of the waveguide. The cross-phase 
modulation parameter$g_{12}$ in the typical case
is approximately equal to 2, which corresponds to the
phase separation mode.

We would like to remind the assumptions for
the model (1). This system is derived for a weakly
inhomogeneous medium in which the background
dielectric permittivity is a function of frequency 
$\varepsilon(\omega)$, so that the corresponding dispersion
law has the following form
$$
k(\omega)=\sqrt{\varepsilon(\omega)}\omega/c. 
$$

The range of the anomalous group velocity dispersion 
of interest to us (where $k''(\omega)<0$ ) is usually near
the low-frequency edge of the transparency window
(in real substances this is often the infrared region
of the spectrum; see, e.g., [46-48]). It follows from
Maxwell's equations that in an optical wave the electric 
field ${\vec E}$ and the electric induction field ${\vec D}$
are related by equation
\begin{equation}
\mbox{rot}\,\mbox{rot}\,{\vec E}=
-\frac{1}{c^2}\frac{\partial^2 {\vec D}}{\partial t^2}.
\label{wave_eq}
\end{equation}
Slow complex envelopes ${\bf E}$ and ${\bf D}$ are introduced
by substitutions
\begin{eqnarray}
&&{\vec E}\approx\mbox{Re}[{\bf E}\exp(ik_0\zeta-i\omega_0 t)],\nonumber\\
&&{\vec D}\approx\mbox{Re}[{\bf D}\exp(ik_0\zeta-i\omega_0 t)],\nonumber
\end{eqnarray}
where $k_0=k(\omega_0)$ is the carrier wave number. We assume 
a weakly nonlinear material dependence of the Kerr type
\begin{eqnarray}
{\bf D} &\approx& \varepsilon(\omega_0+i\partial_t){\bf E}
+\tilde\varepsilon(x,y){\bf E}
\nonumber\\
&+&\alpha(\omega_0)|{\bf E}|^2  {\bf E}
  +\beta(\omega_0)({\bf E}\cdot{\bf E}) {\bf E}^*,
\end{eqnarray}
with operator $\varepsilon(\omega_0+i\partial_t)$, with spatial 
inhomogeneity $\tilde\varepsilon(x,y)$, and with negative functions
$\alpha$ and $\beta$ in the defocusing case. Substitution into (2) 
is then performed, with the small divergence of the electric field 
neglected. The linear operator 
$[k_0-i\partial_\zeta]^2-[k(\omega_0+i\partial_t)]^2$
arising in the course of transformations is represented
as a product of sum and difference, after which the expansion 
by powers of the operator $\partial_t$ leads to an equation 
of the following form
\begin{equation}
2k_0[-i\partial_\zeta-ik_0'\partial_t+k_0''\partial_t^2/2]{\bf E}=s\{{\bf E}\},
\end{equation}
with a right-hand side that includes transverse Laplacian, 
nonlinearity and spatial inhomogeneity. At the end of 
the derivation, to pass to scalar functions $A_{1,2}$,
a substitution is made to
\begin{equation}
{\bf E}\approx \big[({\bf e}_x+i{\bf e}_y) A_1 
    + ({\bf e}_x-i{\bf e}_y) A_2 \big]/\sqrt{2}.
\end{equation}

\begin{figure}
\begin{center} 
\epsfig{file=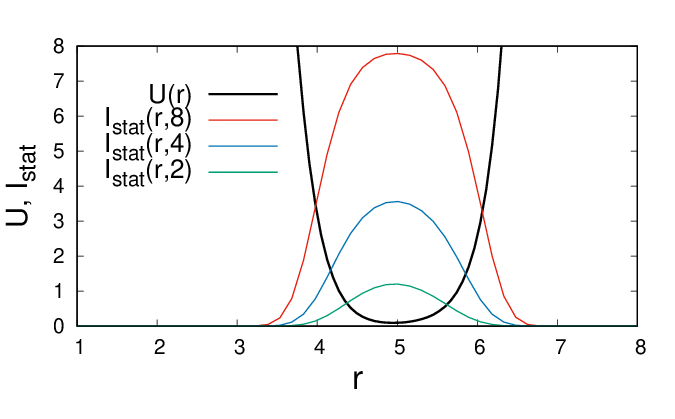, width=80mm}
\end{center}
\caption{
A smooth potential $U(r)$ convenient for numerical modeling, 
roughly corresponding to an annular cross-section waveguide
with a flat bottom and sharp walls. Also, equilibrium wave 
intensity profiles for values of the ``chemical potential'' 
$\mu=2,4,8$ are shown.
}
\label{U_I123} 
\end{figure}

The scale for the transverse coordinates is chosen
to be some large parameter $R_0$. Let it be of the order
of several tens of wavelengths, i.e., up to one hundred 
micrometers. Then the width of the domain wall with 
the normal across the beam at the wave intensities 
of interest $I_{1,2}\lesssim 10$ will be at least a dozen
wavelengths. The natural scale for the variable $\tau$ is the
combination $k_0R_0\delta/\omega_0$, where the dimensionless
parameter
$$
\delta=\omega_0\sqrt{|k_0''|/k_0}
$$ 
characterizes the relative magnitude of the group velocity 
dispersion. Related to this combination is the condition 
$k_0R_0\delta\gg 2\pi$, which ensures quasi-monochromaticity
of the wave on the variable $\tau$. This requirement turns 
out to be quite onerous for small $\delta$,
but we will assume that at least $\delta\gtrsim 0.3$, 
and then there will be at least three wave periods per 
domain wall width with normal along the beam. The longitudinal
coordinate $\zeta$ is measured in units of $k_0 R_0^2$ 
(a few centimeters), and the electric field is measured in units of
$$
\sqrt{2\varepsilon(\omega_0)/|\alpha(\omega_0)|}/(k_0 R_0).
$$
The external potential then is
$$
U=-k_0^2 R_0^2\tilde\varepsilon/2\varepsilon(\omega_0).
$$
As a result, one obtains the non-dimensional equations (1), whereby
$$
g_{12}=1+2\beta(\omega_0)/\alpha(\omega_0).
$$

The well-proven split-step Fourier method of the second order
of approximation for variable $\zeta$ 
was used for numerical modeling. As a preliminary step,
unwanted hard degrees of freedom were ``extinguished'' 
by a dissipative ``imaginary time propagation'' procedure. 
The details of this approach are described in [38-40]. 
The computational domain had dimensions $(6\pi)\times(6\pi)\times T_\tau$,
with periodic boundary conditions, and the period $T_\tau$ 
on the variable $\tau$ was taken equal to $24\pi$ or $48\pi$ 
for different numerical experiments.

\begin{figure}
\begin{center} 
\epsfig{file=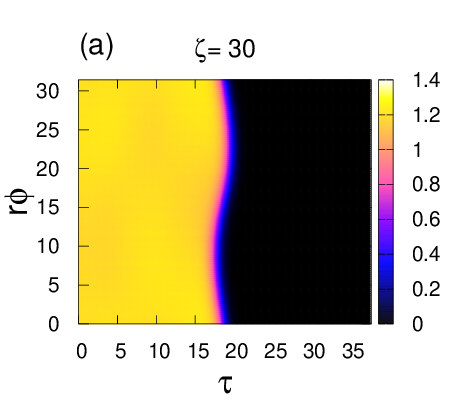, width=42mm}
\epsfig{file=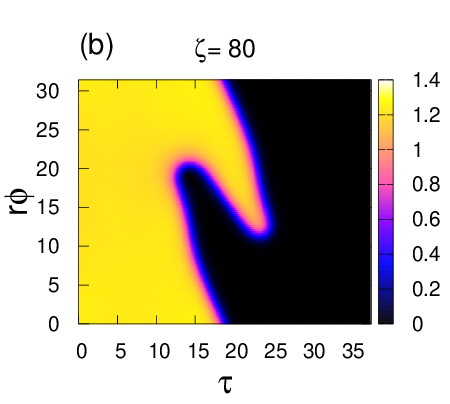, width=42mm}\\
\epsfig{file=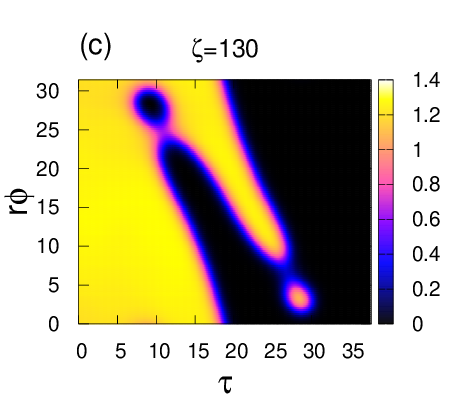, width=42mm}
\epsfig{file=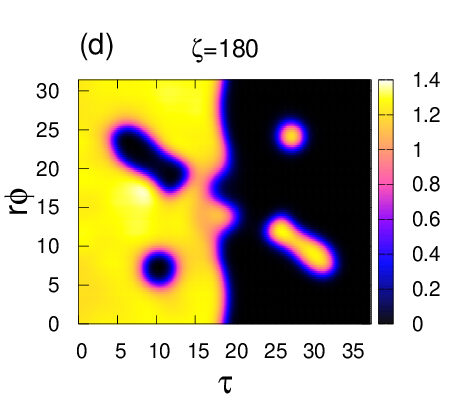, width=42mm}
\end{center}
\caption{Example of the development of Kelvin-Helmholtz instability at $Q=2$
and a small level of nonlinearity $\mu=2$. The intensity of the
first component $I_1$ on a cylinder of radius $r=5$ is shown for
several values of the variable $\zeta$.  
}
\label{I_mu2} 
\end{figure}

\begin{figure}
\begin{center} 
\epsfig{file=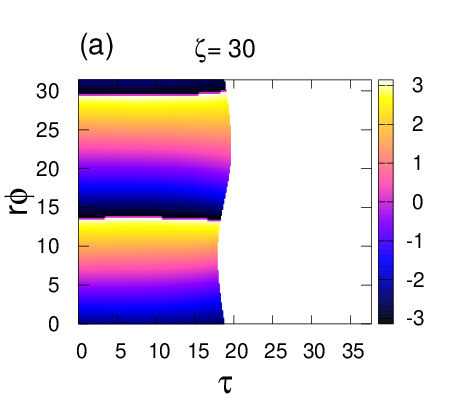, width=42mm}
\epsfig{file=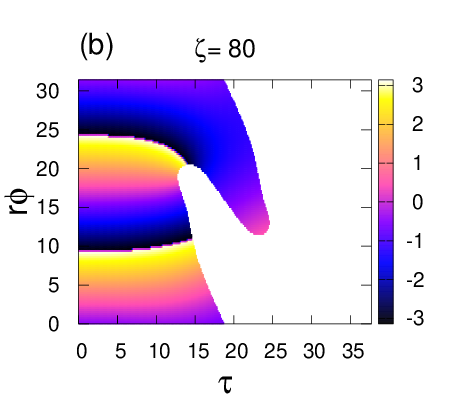, width=42mm}\\
\epsfig{file=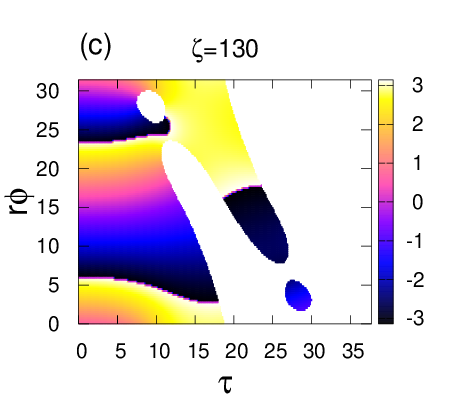, width=42mm}
\epsfig{file=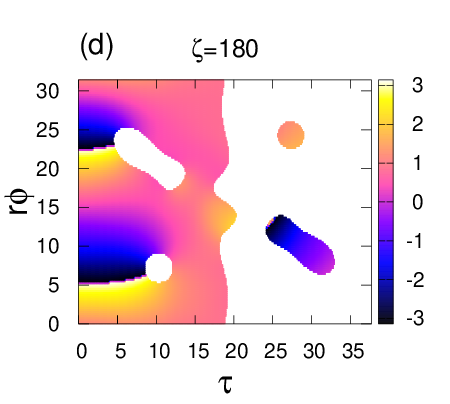, width=42mm}
\end{center}
\caption{Phase $\Theta_1$ of the first components on a cylinder 
of radius $r=5$, corresponding to Fig. 2. Each panel shows
only that part of the cylinder involute where $I_1>0.4$.
}
\label{Theta_mu2} 
\end{figure}

\section{GEOMETRY OF THE PROBLEM}

Let us now turn to the geometrical characteristics
of the model. In contrast to the recent work [38], where
parabolic potentials were considered
$
U=(x^2+\kappa^2y^2)/2,
$
here we focus on annular waveguides with the ``flat''
bottom and sharp walls. Such waveguides are easier 
to produce in practice, since the medium in them
is homogeneous. But for numerical modeling by the
chosen method (with a reasonable spatial grid step)
only sufficiently smooth potentials are suitable. 
Therefore, we will approximate the corresponding deep
rectangular pit only in a qualitative sense, using the
following expression (see the corresponding graph
in Fig. 1):
\begin{eqnarray}
U(r)&=& 27[2+\tanh(1.5(r^2/9-5.0))\nonumber\\
    && \qquad  -\tanh(2.4(r^2/9-1.2))].
\end{eqnarray}

The large common multiplier here is taken quite arbitrarily,
as long as it does not require too small a step
on the variable $\zeta$. As a result of this choice, the average
waveguide radius roughly corresponds to $r=5.0$. This
value will be used in the following to visualize the wave
pattern on a ``median'' cylindrical surface in $(x,y,\tau)$
space at different values of propagation distance $\zeta$.

The initial conditions in our numerical experiments
set the location of the domain wall approximately
across the waveguide axis (its normal is approximately 
along the axis). In this case, the ``currents'' of both
polarizations had an azimuthal character (with coordinate
multipliers $\exp(iQ_{1,2}\phi)$, where $\phi$ is the azimuthal
angle, $Q_1$ and $Q_2\neq Q_1$ are integer ``quantum'' numbers).
Thus, the system is three-dimensional, and hence the
anomalousness of the dispersion is fundamentally important. 
But the difference between the outer $r+a/2$
and inner $r-a/2$ radii of the waveguide is relatively
small (effectively $a/r \approx 0.2$--$0.4$ depending on the
``chemical potential'' of the wave, as can be seen from
the intensity profiles in Fig. 1). In such a geometry, the
system behaves approximately as a spatially two-dimensional 
system (on a cylinder). This contributes
to the manifestation of instability of the type under
consideration, and is also important for the visualization 
of numerical results (see Fig. 2--6).

\begin{figure}
\begin{center} 
\epsfig{file=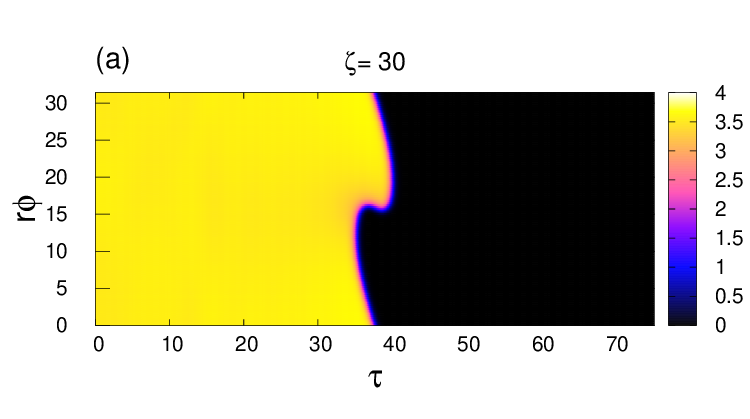, width=77mm}\\
\epsfig{file=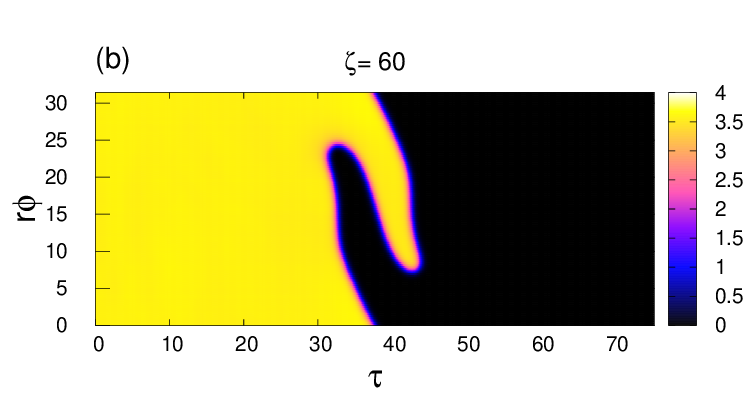, width=77mm}\\
\epsfig{file=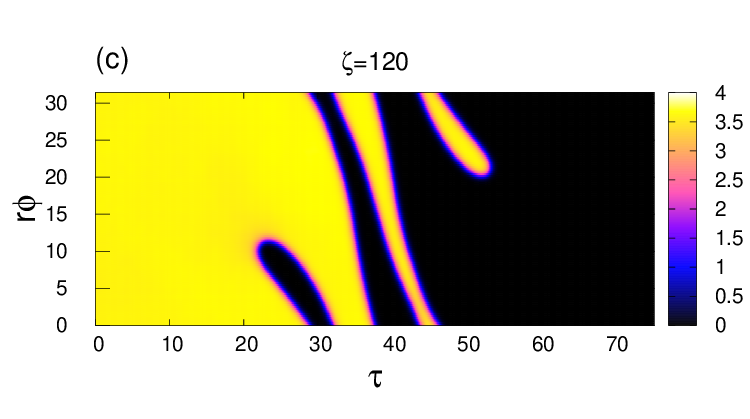, width=77mm}\\
\epsfig{file=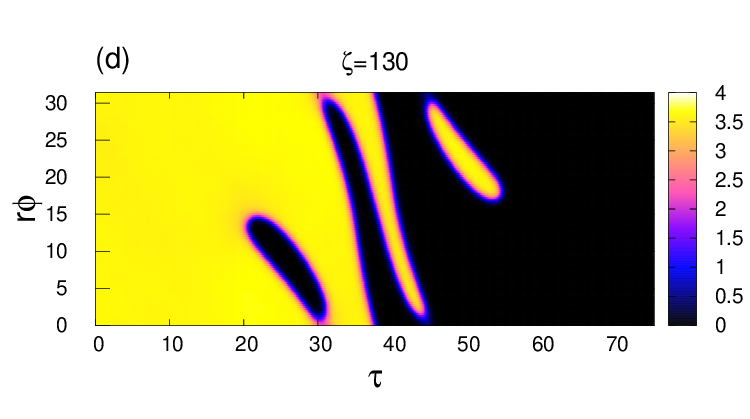, width=77mm}
\end{center}
\caption{Formation of long ``fingers'' at the intermediate stage of
instability at $Q=3$, $\mu=4$.
}
\label{I_mu4_Q3} 
\end{figure}

It should be noted that the system (1) also has solutions 
independent of $\tau$, and there the flow occurs in the
$(x,y)$ plane. The normal to the domain wall also lies
in this plane. On the one hand, these solutions are
attractive because for such stationary configurations
in physical space the requirement of anomalous dispersion 
is removed. And it is quite possible that three-dimensional 
perturbations would not manifest themselves 
at a sufficiently long propagation distance. But,
on the other hand, when the transverse dimensions
of the sample are finite, it becomes necessary to loop
both stationary currents somehow, and then the problem 
of the appearance of additional transverse instabilities 
due to the interaction with transverse boundaries
may arise. There is also a third variant of stationary
flows with a shift along the contact boundary of light
``liquids'' -- when the motion with different velocities
occurs along the axis of the waveguide. This variant
is described by the multipliers $\exp(iv_{1,2}\tau)$ and 
thus corresponds to some difference in the average frequency
of the left and right polarizations. For this configuration, 
the waveguide cross-section need not be circular.
The first and third options seem equally promising.
In this paper, we focus on azimuthal currents.

\begin{figure}
\begin{center} 
\epsfig{file=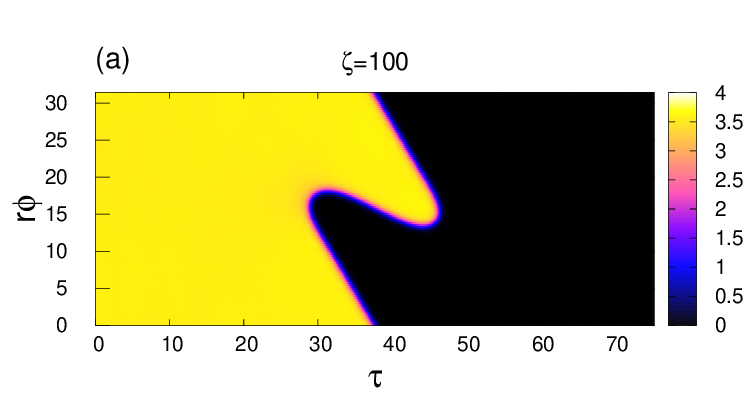, width=77mm}\\
\epsfig{file=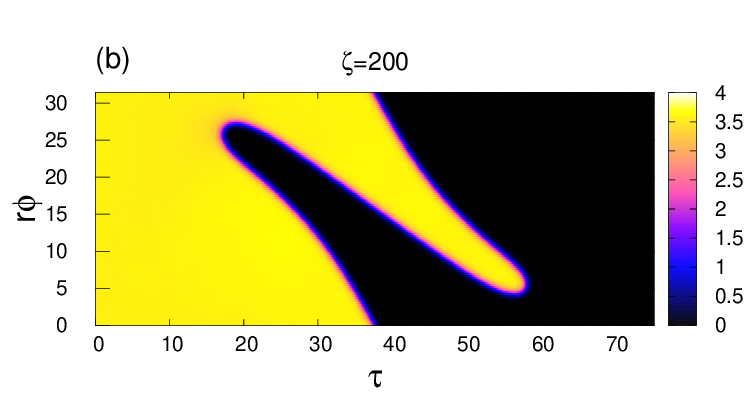, width=77mm}\\
\epsfig{file=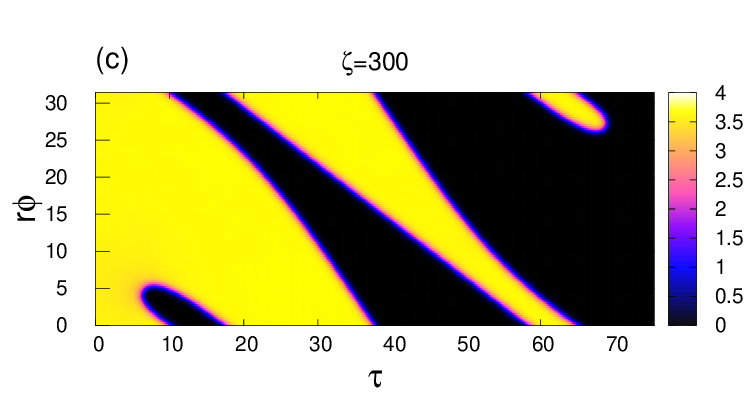, width=77mm}
\end{center}
\caption{Greater slope of the ``fingers'' at a lower relative motion velocity than
in Figure 4 ($Q=2$, $\mu=4$).
}
\label{I_mu4_Q2} 
\end{figure}

The classical formula states that the Kelvin-Helmholtz 
instability increment equals
$$
\Gamma(k)=k\sqrt{v^2/4-(\sigma/2\rho) k},
$$
where $k$ is the wave number, $\sigma$ is the surface tension
coefficient, $\rho$ is the density of the fluid equal on both
sides of the interface, $v$ is the velocity difference for
undisturbed flow. From this formula we can extract
useful information for our case as well. Indeed, at sufficiently 
large values of the ``chemical potential'' $\mu$, the
density of ``liquids'' $\rho=I\approx\mu$, the effective surface
tension coefficient of the domain wall can be estimated 
as $\sigma\propto \mu^{3/2}$ [28], the velocity of relative motion
$v\approx(Q_1-Q_2)/r$, and the basic wave number $k=1/r$.
As a result, we have an estimation formula
\begin{equation}
\Gamma\approx\frac{1}{r} 
\sqrt{\frac{(Q_1-Q_2)^2}{4r^2}-C\frac{\sqrt{\mu}}{r} },
\label{Gamma}
\end{equation}
where $C$ is a coefficient of the order of one. The applicability 
of this formula also requires that the relative
velocity $v$ is small compared to the ``sound speed''
$s\sim\sqrt{\mu}$. If this condition is met, the development
of instability should follow the scenario determined
by the ratio of the two summands under the root sign
(implying a sufficiently large longitudinal interval $T_\tau/2$
between neighboring domain walls). It follows from
formula (7), in particular, that at fixed $r$ and $(Q_1-Q_2)$
a significant increase in the wave intensity should lead
to a complete suppression of instability. This effect
was indeed observed in numerical experiments. For
example, with $Q_1=-Q_2=2$ at $\mu=8$ and $\mu=10$ the
instability developed, whereas at $\mu=12$ it was already absent.

\section{NUMERICAL EXAMPLES}

\begin{figure}
\begin{center} 
\epsfig{file=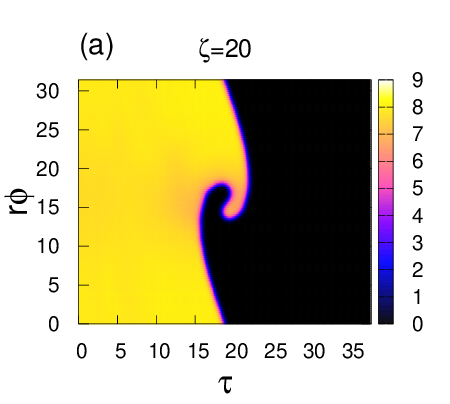, width=42mm}
\epsfig{file=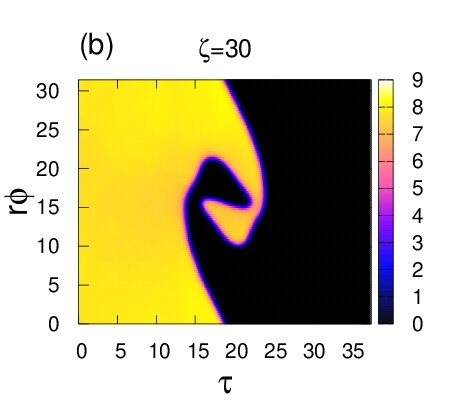, width=42mm}\\
\epsfig{file=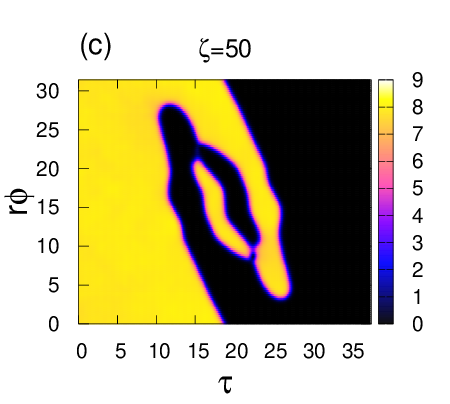, width=42mm}
\epsfig{file=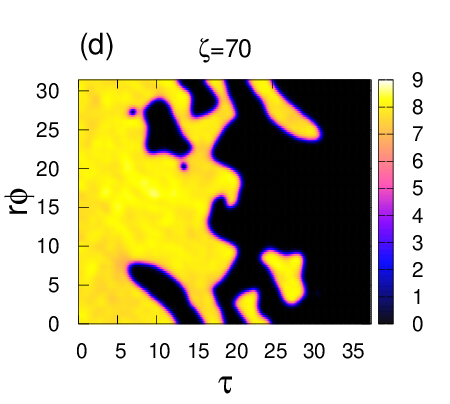, width=42mm}
\end{center}
\caption{Example of the development of Kelvin-Helmholtz instability with 
the formation of a spiral structure at the intermediate stage at 
$Q=5$, $\mu=8$.
}
\label{I_mu8} 
\end{figure}

We should be said that each stationary configuration
on the variable $\zeta$ is characterized by two ``chemical 
potentials'', $\mu_1$  and $\mu_2$. In this case 
$$
A_{1,2}=\sqrt{I_{1,2}(x,y,\tau)}\exp(i\Theta_{1,2}(x,y,\tau)-i\mu_{1,2}\zeta).
$$
In our case the phases $\Theta_{1,2}=Q_{1,2}\phi$. Several examples 
of equilibrium profiles of intensity $I_{\rm stat}(r,\mu)$ in the
presence of only one stationary component are presented in Fig.1.

For simplicity, all numerical experiments were
performed with symmetric initial conditions, so that
$Q_1=-Q_2=Q$, and $\mu_1=\mu_2=\mu$. Therefore, each
panel of Fig. 2-6 shows only half of the cylinder involute. 
In particular, Fig. 2 shows how, for $Q=2$, $\mu=2$,
a small sinusoidal perturbation of the interface as a result
of instability is transformed into a structure shaped like the
silhouette of a human finger. A blob is then detached from
the tip of the ``finger'', within which a vortex ``sits'', as follows
from Fig. 3, which shows the corresponding phase distributions.
Still further in each component is separated also the second similar drop.
Fig. 4 demonstrates that the ``finger'' can extend quite a bit before it
breaks and the drop separates. The angle of inclination of such
``fingers'' on the cylinder with respect to the azimuthal direction
is greater the smaller the relative velocity (parameter $Q$), as can 
be seen by comparing Fig. 5 with Fig. 4. Finally, Fig. 6 shows that 
at high relative velocity ($Q=5$) and wave intensity ($\mu=8$)
at the beginning of the nonlinear instability stage, the boundary
tends to twist into a spiral structure, i.e., the instability
develops according to an almost classical scenario. But
when the spatial scale of the order of the domain wall
thickness $w\sim 1/\sqrt{\mu}$ reached, the dynamics switches
to the ``quantum'' regime with the separation of drops
and quantized vortices.

\section{CONCLUSION}

Thus, this paper theoretically substantiates the fundamental 
possibility of observing Kelvin-Helmholtz instability 
in a defocusing Kerr optical medium with the anomalous 
velocity dispersion. The three-dimensional numerical 
simulation yielded configurations qualitatively 
similar to two-dimensional structures obtained
in [12-14] for an idealized model of a binary BEC
in unbounded space. Obviously, our three-dimensional
results equally apply to BECs placed in a cylindrical
confinement potential with a deep minimum at a finite 
value of the radial coordinate. But for cold gases
such potentials remain unrealistic for the time being,
whereas from the point of view of optics,
there is nothing unusual in a time-length light beam
propagating along a waveguide of a given cross section.
From the practical point of view, the advantage of the
optical structures considered here are also their not 
small spatial dimensions in comparison with BECs, 
not to mention the absence of extreme temperature 
constraints.

We used the simplest model, which does not take
into account dispersion corrections of higher orders.
In reality, the parameter $\delta$ can be so small that (at not
too large values of the product  $k_0 R_0$) deviations from
the quadratic dispersion will have a noticeable effect.
However, as shown by additional numerical experiments 
(their results are not presented here), taking into
account the relatively small third-order dispersion,
although it distorts the picture of the instability development 
somewhat, but does not spoil it qualitatively
at a sufficiently large distance.

\vspace{2mm}

{\bf FUNDING}.
The work was carried out within the framework
of the State Assignment No. 0029-2021-0003.

\end{document}